\documentclass[aps,prl, onecolumn,showpacs,floatfix,superscriptaddress]{revtex4}

\usepackage{graphicx}
\usepackage{dcolumn}
\usepackage{amsmath}
\usepackage{amssymb}
\usepackage{dsfont}

\newcommand{\be}{\begin{equation}}
\newcommand{\ee}{\end{equation}}
\newcommand{\fig}[1]{Fig.~\ref{#1}}
\newcommand{\eq}[1]{Eq.~(\ref{#1})}

\begin{document}

\title{NMR experiment factors numbers with Gauss sums}

%
%
%
%

\author{Michael Mehring\email[correspondence to:]{m.mehring@physik.uni-stuttgart.de}}
\affiliation{Physikalisches~Institut, Universit\"at~Stuttgart,
D-70550 Stuttgart, Germany}
\author{Klaus M\"uller}
\affiliation{Physikalische Chemie, Universit\"at~Stuttgart,
D-70550 Stuttgart, Germany}
\author{Ilya {Sh.} Averbukh}
\affiliation{Department of Chemical Physics, Weizmann Institute of Science, Rehovot 76100, Israel}
\author{Wolfgang Merkel}
\author{Wolfgang P. Schleich}
\affiliation{Abteilung f\"ur Quantenphysik, Universit\"at~Ulm,
D-89081 Ulm, Germany}

\date{\today}
\begin{abstract}\centerline{
We factor the number 157573 using an NMR
implementation of Gauss sums.}
\end{abstract}
\maketitle
According to the legend\cite{derbyshire:2003} Karl Friedrich Gauss as a child was given by his teacher the problem of adding all integers up to a given number $m$. His elegant solution rearranges  the terms of the series in a convenient way arriving at the result
\be
\sum_{k=1}^m k= \frac{1}{2}m(m+1)
\label{young:gauss}
\ee
which depends quadratically on $m$. In his adult life Gauss revisited square dependencies in the context of number theory\cite{davenport:1980} when he calculated the sum of quadratic  phase factors which today carry the name Gauss sums\cite{schleich:2005:primes}.  In this contribution we present the first implementation of a factorization algorithm \cite{merkel:fortschritte:2006,merkel:PRA:part1}  based on Gauss sums\cite{footnote:van:dam} by factorizing the number $N =157573$ using NMR techniques.
It is intriguing that the physical realization relies on the sum formula \eq{young:gauss}.

Factorization of numbers into their prime factors is considered a
hard non-polynomial problem\cite{stenholm:quantumapproach:2005} for classical computers. It was Shor \cite{shor:1994} who proposed a quantum algorithm which can
solve the problem on a quantum computer\cite{nielsen:chuang:2000} with a tremendous speedup as compared to a classical computer. Vandersypen et al.\cite{chuang:nature:2001} demonstrated in an NMR implementation of the Shor algorithm with seven qubits the factorization of the number $15 = 3\cdot5$.

Gauss sums are the discrete version of the Fresnel integrals\cite{born:wolf} familiar from classical optics effects such as diffraction from a wedge or the Feynman formulation of quantum mechanics. Thus they are a sum rather than an integral over quadratic phase factors. Due to this discreteness Gauss sums have interesting periodicity properties which manifest themselves in the Talbot effect\cite{schleich:2001}, fractional revivals\cite{schleich:2001:thebook} or curlicues\cite{berry:curlicues}. Due to this periodicity property they not only play an important role in number theory but are also an ideal tool to factor numbers. Indeed several such Gauss sum factorization algorithms\cite{clauser:1996,harter:2001,mack:2002,merkel:ijmpb:2006,merkel:fortschritte:2006,merkel:PRA:part1} have been put forward. All these schemes capitalize on the periodicity properties but differ in their method of implementation ranging from a $N$-slit Young interferometer\cite{clauser:1996} via molecules\cite{harter:2001} and wave packet dynamics\cite{mack:2002} to chirped laser pulses interacting with atoms\cite{merkel:ijmpb:2006,merkel:fortschritte:2006,merkel:PRA:part1}. However, so far no experimental demonstration of this approach has been provided\cite{footnote0}.

In the present paper we use NMR techniques to realize a rather special Gauss sum which brings out factors with a remarkable contrast even when only a few terms appear in the sum. We start by briefly summarizing the essential ingredients of our scheme. Since we do not employ entanglement yet, we still need exponential resources. Nevertheless, the quasi-random interference\cite{footnote3} contained in the Gauss sums allows us to work with only a few terms in the sum. In the experiment this advantageous feature translates into the need for a few pulses within a decay time of the system only. We then turn to the experiment and demonstrate the power of the scheme for a number with six digits, but emphasize that extensions to much larger numbers are within reach.

Our experiment implements the Gauss sum\cite{footnote1a}
\be
{\cal A}_N^{(M)}(\ell) = \frac{1}{M+1} \sum_{m=0}^M \exp\left[- 2\pi i\, m^2\frac{N}{\ell}\right],
\label{eqn:Gauss}
\ee
with $M+1$ terms and $N$ is the number to be factored. The argument $\ell$ with $1\leq\ell \leq
\sqrt{N}$ scans through all integers between $1$ and $\sqrt{N}$ for possible
factors.

\begin{figure}[htbp]
\centering
\includegraphics[width=0.65\textwidth]{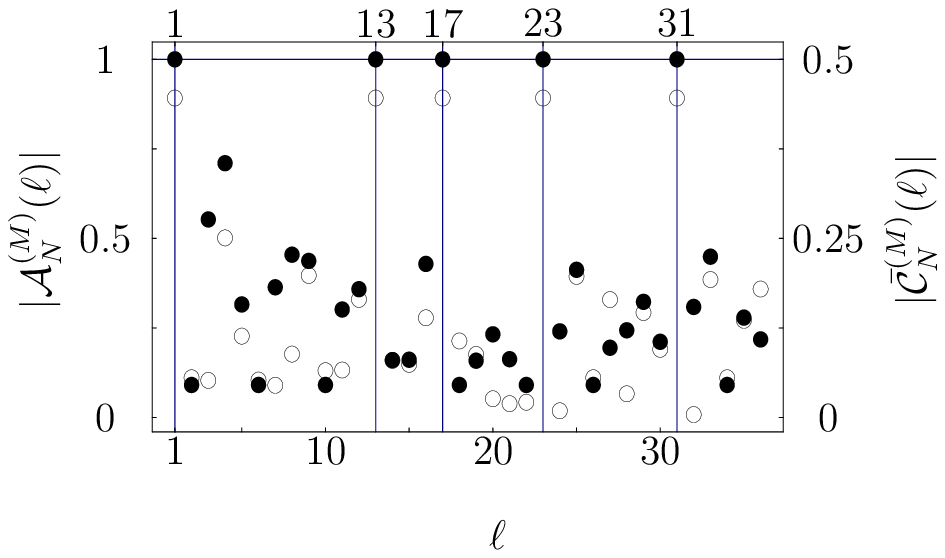}
\caption{Factorization interference pattern for $N=157573=13 \cdot 17 \cdot 23 \cdot31$ obtained from the Gauss sum
${\cal A}_N^{(M)}$ (black dots) and from the Gauss sum $\bar{{\cal C}}_N^{(M)}$  corresponding to damping (circles). In
both cases already $M+1=11$ terms are sufficient to clearly discriminate the factors from non-factors. Note the
different scales for ${\cal A}_N^{(M)}$ and ${\cal C}_N^{(M)}$.} \label{A157573}
\end{figure}

The capability of the Gauss sum, \eq{eqn:Gauss}, to factor numbers stems from the fact that for an integer factor $q$ of $N$ with $N=q \cdot r$ all phases in ${\cal A}_N^{(M)}$ are integer multiples of $2\pi$. Consequently the terms add up constructively and yield ${\cal A}_N^{(M)}(q)=1$ as shown in \fig{A157573} by black dots.
When $\ell$ is not a factor the quadratic phases oscillate rapidly with $m$ and ${\cal A}_N^{(M)}$ takes on small values.

\begin{figure}[htbp]
\centering
\includegraphics[width=0.55\textwidth]{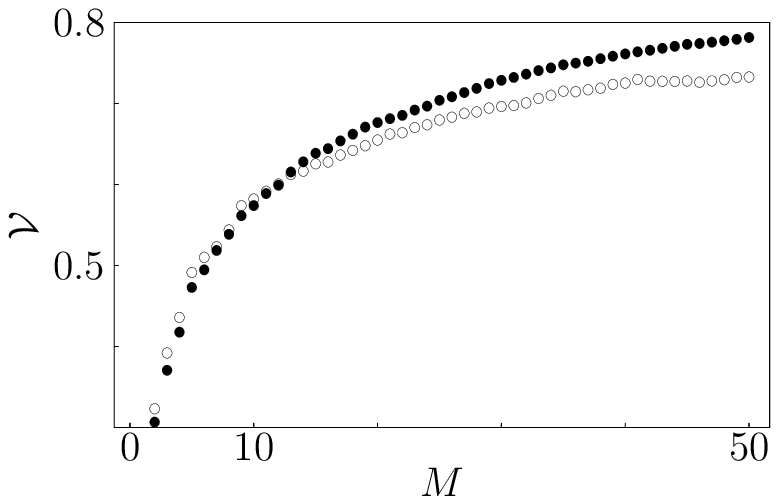}
\caption{The contrast ${\cal V}$ of the factorization interference pattern of \fig{A157573} as a function of the number $M$ of terms in the Gauss sum ${\cal A}_N^{(M)}$ for $N=157573$ (circles) and $N=4683359$ (black dots).}
\label{contrast}
\end{figure}

In this process of destructive interference the truncation parameter $M$ plays a crucial role. Indeed, already a few terms in the sum are sufficient to discriminate factors from non-factors in the signal $|{\cal A}_N^{(M)}(\ell)|$.
In order to analyze the dependence of this surprising feature on $M$ we define the contrast\cite{born:wolf}
${\cal V}\equiv (1-a)/(1+a)$
of the factorization interference pattern in complete analogy to classical optics where
$a \equiv \sum_{\ell'=1}^{n_0} |A_N^{(M)}(\ell')|/n_0$ denotes the average value of the sum at the non-factors upto $\sqrt{N}$. Indeed, $n_0$ is the closest integer to $\sqrt{N}$ and the summation runs over all arguments $\ell'$ which are not factors of $N$. In \fig{contrast} we show ${\cal V}$ for different values of $N$.
Already a relatively small number of terms $M$ results in good contrast of the signal.

Next we address the complexity of our factorization scheme.
To gain information on the factors of $N$ we have to measure the signal $|{\cal A}_N^{(M)} (\ell)|$ for arguments $\ell$ belonging to the interval $[0,n_0]$.
Since at most $\sqrt{N}$ data-points $\{\ell,|{\cal A}_N^{(M)}(\ell)|\}$ have to be acquired, we estimate the required resources as
$
\sqrt{N}=\exp\left[L/2\right]
$
where $L=\log N$ is the number of digits of $N$. Although our scheme scales exponentially we profit from the small number of terms $M$ necessary to distill the factors.

\begin{figure}[htbp]
\centering
\includegraphics[width=0.65\textwidth]{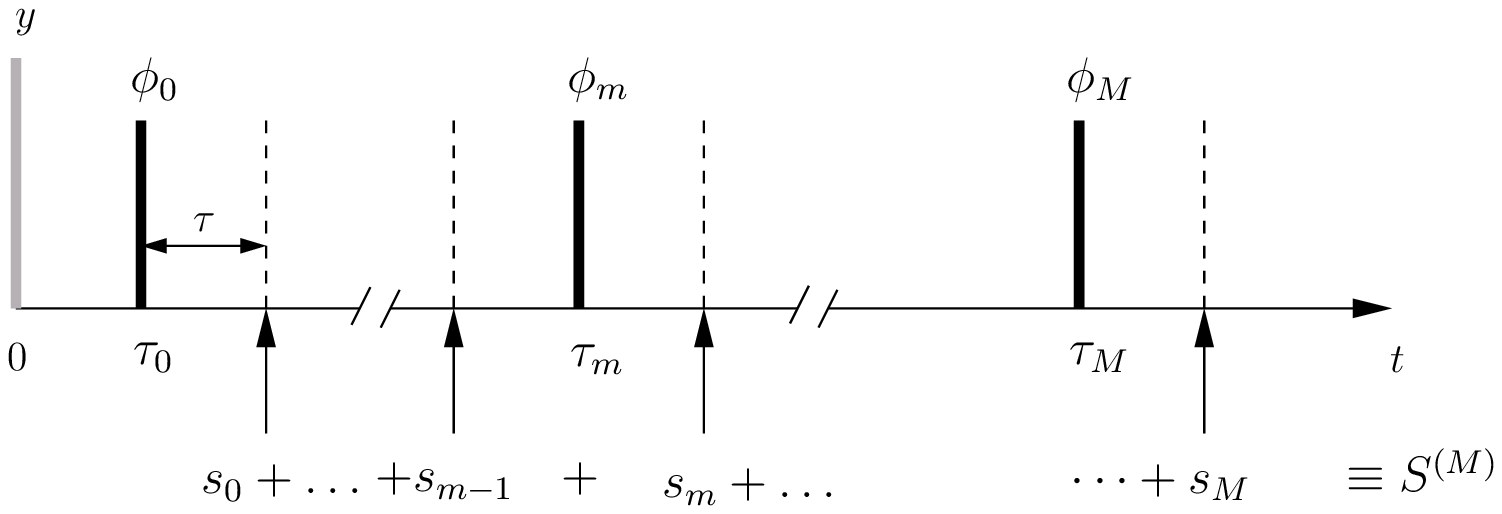}
\caption{NMR implementation of our factorization scheme using Gauss sums.
The $y$-pulse which prepares the initial density matrix $\rho_{\rm in}$ is followed after a time $\tau$ by a pulse which imprints the phase $\phi_0$ on the spins. This pulse is the first of a sequence of $M+1$ pulses each imprinting a different phase $\phi_m$. At times $\tau_m+\tau\equiv(2m+1)\tau+\tau$ we measure the echo, that is the polarization $s_m$ in $x$-direction and sum the echos $s_m$ over all $m$ to obtain $S^{(M)}$. In order to implement a Gauss sum the phases $\phi_m$ need to be proportional to the number $N$ to be factored and have to increase linearly as a function of $m$ since the spin dynamics expressed by $s_m$ depends on the sum over all phases of the previous pulses.}
\label{fig:pulse_seq1}
\end{figure}

We now turn to our experimental realization of a Gauss sum. In the original proposal\cite{merkel:fortschritte:2006} the Gauss sum arises through the time evolution of a two-level atom whose transition frequency increases linearly in time and which is driven by a train of laser pulses. In our experiment we subject an ensemble of spins $I=1/2$ to a specific sequence of RF pulses as shown in \fig{fig:pulse_seq1}.
They lead to a sequence of signals which when summed represent a Gauss sum closely related to \eq{eqn:Gauss}.

In our experiment we use $\text{H}_2\text{O}$ as an ensemble of
protons with nuclear spin 1/2 in Boltzmann equilibrium at room
temperature described by a density operator
$
\rho_B\equiv\mathds{1}/2-\epsilon\,I_z,
$
with the identity operator $\mathds{1}$ and $0<\epsilon\ll1$. Throughout the paper we use the notation $I_j \equiv \sigma_j/2$ where $\sigma_j$ are the familiar Pauli spin matrices.

Radio frequency pulses are applied at the Larmor frequency 400~MHz of the protons. Experiments were performed with a
Varian Infinity Plus NMR spectrometer applying a cycle time of $t_c =2\tau= 100 \mu s$ and with relaxation time $T_2
\approx 200 ms$. The pulse sequence shown in \fig{fig:pulse_seq1} is based on the Carr, Purcell, Meiboom and Gill
(CPMG)-sequence\cite{carr:54,meiboom:58} which was proposed almost 50 years ago for measuring $T_2$ relaxation times in
inhomogeneous fields. For our purpose we have modified this sequence such that the resulting evolution of the proton
spins expressed by a sequence of echos which leads to the desired Gauss sum.

The pulse sequence consists of an initiating $\pi/2$-pulse in
$y$-direction which creates the initial density matrix $ \rho_{\rm
in} \equiv \mathds{1}/2-\epsilon\,I_x. $ This pulse  is followed
by a series of $M+1$ $\pi$-pulses with separation $2\tau$ which
are individually phase shifted with respect to the $x$-axis of the
rotating frame by an angle $\phi_k$\cite{warren:80,ernst:87}.
In the laboratory system the corresponding Hamiltonian reads
\be
H(t)=\hbar\omega_0\, I_z + 2\pi\hbar \sum\limits_{k=0}^M \delta(t-\tau_k) \cos(\omega\tau_k-\phi_k)\,I_x
\label{Hamiltonian}
\ee
where $\omega_0$ and $\omega$ denote the frequencies of the transition and the driving field, respectively. The pulses act at times $\tau_k \equiv (2k+1)\tau$ and the phases $\phi_k$ will be chosen later in  order to obtain a Gauss sum.

With the help of the identity
\be
U_j(\alpha)\equiv\exp(-i\alpha I_j)
=\cos(\alpha/2)\,\mathds{1}-2i\,\sin(\alpha/2)\, I_j
\label{identity}
\ee
the Hamiltonian $\widetilde{H}$ within the rotating wave approximation, that is in the frame rotating with the frequency $\Delta\omega=\omega_0-\omega$ takes the form
\be
\widetilde{H}=\hbar\Delta\omega\, I_z + \pi\hbar \sum\limits_{k=0}^M \delta(t-\tau_k) \left(\cos\phi_k\,I_x+\sin\phi_k\,I_y\right)
\label{Hamiltonian2}
\ee
and the resulting time evolution operator
\be
{\cal U}_k
\equiv
U_z(\Delta\omega\tau)
U_z(\phi_k)U_x(\pi)U_z^\dagger(\phi_k)
U_z(\Delta\omega\tau)
\label{U:MM}
\ee
of the $k$-th cycle reflects the three stages shown in \fig{fig:pulse_seq1}: ({\it i}) Free time evolution for the time $\tau$ followed by ({\it ii}) a $\pi$-pulse which imprints the phase $\phi_k$, and ({\it iii}) another free time evolution during the time $\tau$. The $\pi$-pulse eliminates the effects of the free time evolution and ensures that the decoherence due to an
inhomogeneous distribution of local fields is compensated by refocussing.

At times $\tau_m+\tau$ we measure the polarization
\be
s_m \equiv
\frac{\text{Tr}\left( I_x \rho_m\right)}
{\text{Tr}\left( I_x \rho_{\rm in}\right)}
\label{signal}
\ee
in $x$-direction with
$\rho_m\equiv \rho(\tau_m+\tau)={\cal U}_m \rho_{\rm in}{\cal U}_m^\dagger $ where
${\cal U}_m\equiv {\cal U}(\tau_m+\tau)=\prod_{k=0}^m {\cal U}_k $
is the product of the time evolution operators ${\cal U}_k$ due to all previous pulses.

When we apply \eq{identity} to \eq{U:MM} we find
\be
{\cal U}_k=(-i)
\begin{pmatrix}
0&{\rm e}^{-i\,\phi_k}\\
{\rm e}^{i\,\phi_k}&0
\end{pmatrix}
\label{unitary}
\ee
which yields for the sum
$
S^{(M)}\equiv \sum_{m=0}^M s_m
$
over the signals $s_m$ given by \eq{signal} the expression
\be
S^{(M)}=\sum\limits_{m=0}^M \cos\left(\sum\limits_{k=0}^m (-1)^k 2\phi_k\right).
\label{sum:over:cos}
\ee
Hence, the spin dynamics expressed by the signal $s_m$ depends on its complete
history, that is the phases of all previous pulses. In particular, they enter as an alternating sum. It is here that the sum \eq{young:gauss} of the young Gauss comes into play. When we compare $S^{(M)}$ to the Gauss sum of \eq{eqn:Gauss} we recognize that for the choice
\be
\phi_k=
\left\{
\begin{array}{cc}
(-1)^k(2k-1) \pi \frac{N}{\ell} &\text{ for } k \ge 1\\[2mm]
0&\text{ for } k =0
\end{array}
\right.
\label{phases}
\ee
of the phases $\phi_k$ \eq{sum:over:cos} takes the form
\be
\frac{1}{M+1}S^{(M)}=
\frac{1}{M+1} \sum_{m=0}^M\cos\left(2\pi\, m^2\frac{N}{\ell}\right)\\
\equiv {\cal C}_N^{(M)}(\ell)
\ee
with ${\cal C}_N^{(M)} = \text{Re}\,{\cal A}_N^{(M)}$.

\begin{figure}[htbp]
\centering
\includegraphics[width=0.55\textwidth]{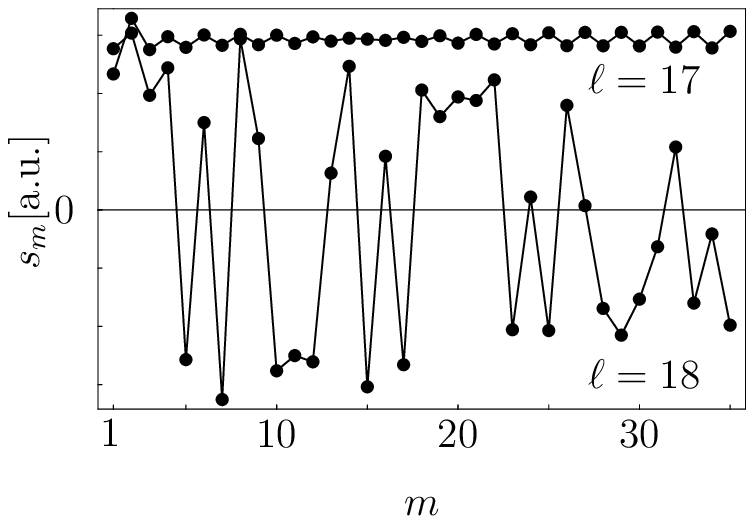}\\[5mm]
\includegraphics[width=0.55\textwidth]{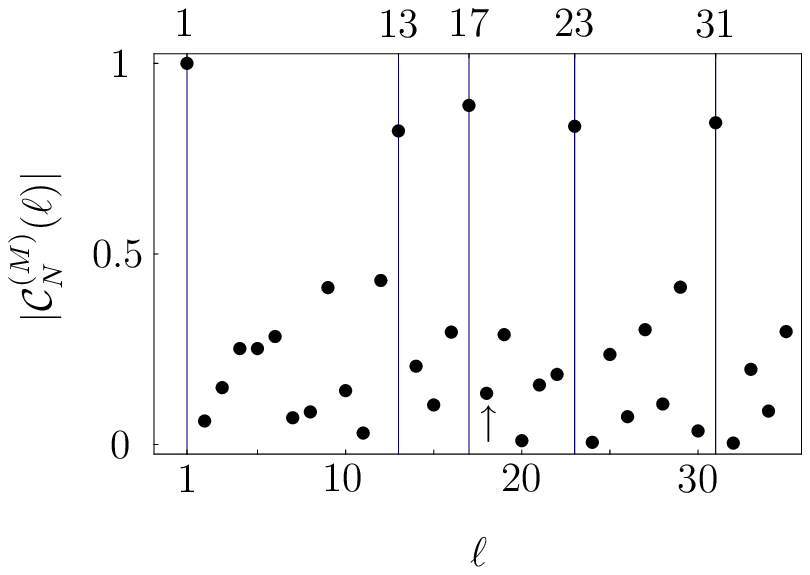}
\caption{Experimental realization of factoring $N = 157573=13 \cdot 17 \cdot 23 \cdot 31$ using the NMR implementation of the Gauss sum ${\cal C}_N^{(M)}$.
In the top and bottom we depict the echo height $s_m$ measured at times $\tau_m+\tau$ and the resulting average ${\cal C}_N^{(M)}$ for different trial factors $\ell$, respectively. For factors such as  $\ell=17$ the signals $s_m$ are approximately constant as a function of $m$ with an average value ${\cal C}_N^{(M)}(17)$ close to unity. In contrast, for a non-factor such as $\ell=18$ $s_m$ oscillates around zero and when summed over $m$ almost averages out as indicated by the arrow.}
\label{fig:plot157573}
\end{figure}
In \fig{fig:plot157573} we display the results of our NMR implementation of our factorization scheme based on Gauss sums for $N =157573 =13\cdot 17\cdot
23\cdot 31$. On the top we show the time evolution of the spin under the influence of the particular pulse sequence given by the Hamiltonian $H$ and the phases $\phi_k$ defined by Eqs.~(\ref{Hamiltonian}) and (\ref{phases}), respectively. As a measure of the dynamics we show the echo signal $s_m$ following from \eq{signal}. For factors of $N$ such as $\ell=17$ the signal is constant. Consequently we find for the average ${\cal C}_N^{(M)}(\ell=17)$  a value close to one as indicated in the bottom. However, for non-factors such as $\ell=18$ the echo signal oscillates around 0 and leads to a rather small average value ${\cal C}_N^{(M)}(\ell=18)$, shown by the arrow. We emphasize that due to the quasi-random interference of Gauss sums $M=11$ terms are sufficient to discriminate factors from non-factors.

Although decoherence is not a limiting factor in our scheme it is still interesting to investigate its influence. We take incoherent processes into account phenomenologically by introducing a $T_2$ relaxation process and   \eq{sum:over:cos} is modified
\begin{equation}
\label{eqn:SxMT2fin} \bar{{\cal C}}_N^{(M)}(\ell) = \frac{1}{M+1}\sum_{m=0}^M \exp\left(-m
\frac{2\tau}{T_2}\right)\cos\left(2\pi\, m^2\frac{N}{\ell}\right).
\end{equation}
We find that even for an appreciable decay, that is for $2M\tau/T_2=2$ at the end of the sequence the
pattern shown in \fig{A157573} by circles looks almost identical
to the one without decoherence indicated by black dots, except for
the reduced scale of the signal strength. This result is
surprising since the signal $s_M$ at the end of the sequence has
decayed to 13.5\% of its initial value. Hence, decoherence does
not significantly influence our ability to distinguish factors
from non-factors.

\begin{figure}[htbp]
\centering
\includegraphics[width=0.55\textwidth]{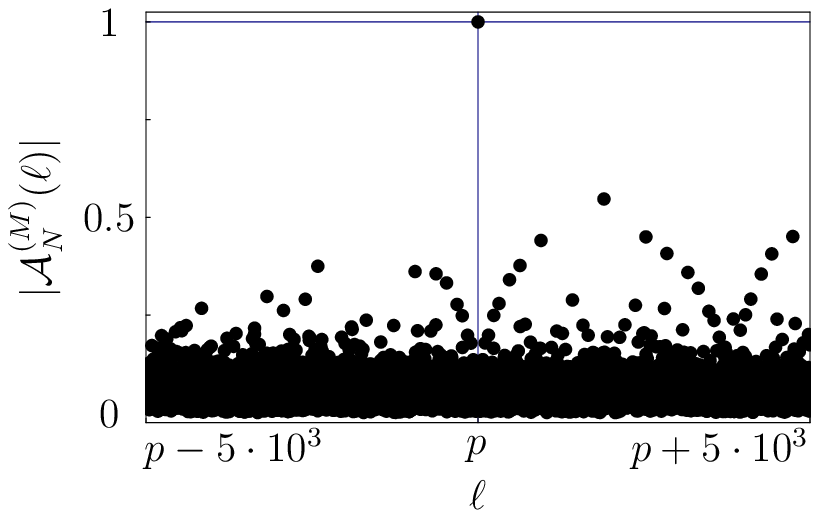}
\caption{Probing the limits of our factorization scheme.
Factorization interference pattern for the 24-digit number $N=1062885837863046188098307=p\cdot q$ obtained from a numerical simulation of the Gauss sum ${\cal A}_N^{(M)}$ in the immediate neighborhood of the prime factor $p=790645490053$ for $M=200$.}
\label{Alimits}
\end{figure}
In summary we have presented the first experimental implementation
of a factorizing algorithm based on Gauss sums with an ensemble of
single qubits.  We have exemplified this technique factoring the
number $N=157573$. However, we claim that extensions to larger numbers are
readily possible  since our method capitalizes on the
quasi-randomness of the phases of Gauss sums. As an outlook we factor in \fig{Alimits} the 24-digit number
$$
N=1062885837863046188098307
$$
with only $M=200$ pulses. Since in NMR it is possible to even have up to $10^4$ pulses within the decay time $T_2$ several new questions emerge:  ({\it i}) What is the optimal number $M$ of terms in the Gauss sum, that is how many pulses are needed for a given $N$ in order to discriminate factors from non-factors? ({\it ii}) How to overcome pulse errors? and ({\it iii}) how to employ entanglement in order to achieve a speed-up?

Obviously, the answers require a more detailed analysis. Therefore we can only speculate. However, our numerical experiments suggest a logarithmic dependence of $M$ on $N$. In addition, pulse error correction techniques based on optimal control theory offer themselves. Finally, the entanglement of two and more spins might open a possibility to reduce the complexity.

We thank B.~Girard, D.~Haase, E.~Lutz, H.~Mack, H.~Meier and G.~G.~Paulus for many fruitful discussions. Moreover, we are grateful to D.~Suter for informing us about his experiment. We appreciate the support of the the Landesstiftung Baden-W\"{u}rttemberg in the framework of the Quantum Information Highway A8. The work of W.~P.~S. is also partially supported by the Max-Planck-Award.


\begin{thebibliography}{24}
\bibitem{footnote0}
{D. Suter has kindly informed us that he has implemented a related, but
different NMR sequence leading to similar results (private
communication).}
\bibitem{footnote3}
{The interference of quadratic phases is at the heart of dynamical localization in the kicked rotor.}
\end{thebibliography}

\end{document}